\documentstyle[twocolumn,epsfig]{jpsj}

\def\runtitle{d-Wave Superconductivity Induced by Chern-Simons Term in
High-$T_c$ Cuprates}
\def\runauthor{Takao {\sc Morinari}}

\title{d-Wave Superconductivity Induced by Chern-Simons Term in
High-$T_c$ Cuprates}

\author{Takao {\sc Morinari}}
\inst{Yukawa Institute for Theoretical Physics, Kyoto University,
Kyoto 606-8502, Japan}

\recdate
{October 5, 2000}

\abst
{
We show that a Chern-Simons term for a gauge field describing a
fluctuation of spins is induced by integrating out hole
fields in the presence of spin-orbit coupling which originates from a
buckling of the CuO$_2$ plane.
Through the Chern-Simons term, holes behave like skyrmion excitations
in a spin system and become a superconducting state with
$d_{x^2-y^2}$ symmetry after the antiferromagnetic long-range order is
destroyed.}

\kword
{high-$T_c$ superconductivity, mechanism of d-wave superconductivity,
Chern-Simons term, skyrmion excitation, buckling of CuO$_2$ plane,
spin-orbit coupling}

\begin{document}
\sloppy
\maketitle
Since the discovery of high-$T_c$ superconductivity in cuprates
\cite{BEDNORZ_MULLER}, much experimental and
theoretical effort has been invested to clarify its mechanism of
superconductivity. 
Results of experimental studies indicate that the following properties
are essential features.
First, the CuO$_2$ layered structure is intrinsic to superconductivity 
and both the undoped and carrier-doped CuO$_2$ planes are characterized
as two-dimensional systems on the basis of their
magnetic\cite{SHIRANE_ETAL}, transport\cite{ITO_ETAL}, and
optical properties\cite{HOMES_ETAL}. 
Second, superconductivity occurs in a disordered spin
background\cite{ANDERSON}.
For the undoped case, the system is a Mott insulator and spins at Cu
sites show antiferromagnetic long-range order below the N{\' e}el
temperature $T_N$. Upon doping, $T_N$ decreases to zero and spin-glass 
behavior is observed\cite{SG}.
Superconductivity emerges upon further doping.
Apparently, disorder in the spin system is introduced by doped
holes.
Third, the Cooper pair is spin-singlet and has $d_{x^2-y^2}$
symmetry.\cite{D_WAVE}

In addition to these properties, the occurrence of superconductivity
appears to be closely related to the structure of the CuO$_2$ plane.
Near the overdoped region of La$_{2-x}$Sr$_x$CuO$_2$ the
disappearance of superconductivity was observed at an orthorhombic
to tetragonal structural phase transition point\cite{TAKAGI_ETAL}.
Moreover, the transition temperature $T_c$ is closely related to
the buckling of the CuO$_2$ plane \cite{CHMAISSEM_ETAL}.
Among the effects on the conduction electron system accompanied by
a buckling of the CuO$_2$ plane, there is spin-orbit
coupling\cite{SPIN_ORBIT1,SPIN_ORBIT2,SPIN_ORBIT3}.
Spin-orbit coupling can have an important effect on
conduction electrons through the Berry phase induced by
the background spin configuration\cite{MN_BERRY}.

\begin{figure}[htbp]
\center
\epsfxsize=2.7truein
\epsfig{file=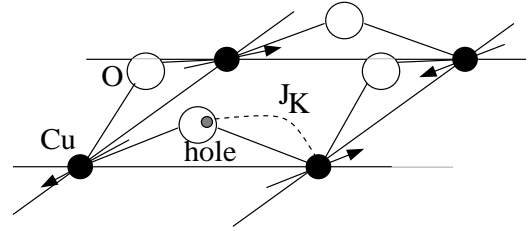,width=2.7in,angle=0}
\vspace{-0.1in}
\caption{Buckling of the CuO$_2$ plane. O atoms are displaced from the
CuO$_2$ plane. Arrows represent the spins at Cu sites. The hole at
the O site interacts with the spin via Kondo coupling $J_K$.}
\label{fig_buckling}
\end{figure}

In this Letter, we propose a mechanism of d-wave superconductivity in
a disordered spin background based on a two-dimensional model with
spin-orbit coupling. 
We assume for the spin-orbit coupling term that it is induced by the
buckling of the CuO$_2$ plane, as shown in Fig. \ref{fig_buckling}.
We show that a Chern-Simons term for a gauge field, which describes
the fluctuation of spins at Cu sites, is induced by integrating out
the hole fields.
Through this Chern-Simons term, holes behave like skyrmion
excitations\cite{SKYRMION1,SKYRMION2,SKYRMION3,SKYRMION4,SKYRMION5,
MARINO_NETO}
for the spin system.
When the antiferromagnetic long-range order is destroyed by these
skyrmion excitations, the Chern-Simons term leads to Cooper pairing of 
holes.
We show that the pairing state is spin-singlet with
$d_{x^2-y^2}$ symmetry using a transformation to the previously
considered model\cite{MORINARI}. 

Our model is described by the following Hamiltonian:
\begin{equation}
H = -t_0 \sum_{\langle i,j \rangle} \left( c_i^{\dagger}
c_j + H.c. \right) + J_{\rm K} \sum_j {\bf s}_j \cdot {\bf
S}_j + H_{\rm so} + H_{\rm spin},
\label{hamiltonian}
\end{equation}
where 
$c_i^{\dagger} = \left( \begin{array}{cc}
c_{i\uparrow}^{\dagger} & c_{i\downarrow}^{\dagger} \end{array} \right)$ 
and $c_i=\left( \begin{array}{cc} c_{i\uparrow} &
c_{i\downarrow} \end{array} \right)^T$ denote a creation and
an annihilation operator for holes in a spinor
representation. In the first term, the summation is taken over the
nearest-neighbor sites. 
The second term denotes Kondo coupling between the spin of holes:
${\bf s}_j=\frac12 c_j^{\dagger}
{\mbox{\boldmath ${\bf \sigma}$}} c_j$, and the spin at Cu sites:
${\bf S}_j$\cite{IMADA_ETAL,MATSUKAWA_FUKUYAMA}.
The Hamiltonian (\ref{hamiltonian}) is based on a model which
distinguishes electrons at the Cu site and holes at the O site.
We implicitly exclude the double occupancy of holes at O sites
because there is a strong on-site Coulomb repulsion.
A similar model without $H_{\rm so}$ is proposed in
refs.~\citen{IMADA_ETAL}, ~\citen{MATSUKAWA_FUKUYAMA}
and ~\citen{KAMIMURA_ETAL}.
Note that
if we take the limit $J_K \rightarrow \infty$, eq. (\ref{hamiltonian}) 
without $H_{\rm so}$ corresponds to
the so-called t-J
model\cite{ZHANG_RICE,MATSUKAWA_FUKUYAMA}.
For spin orbit coupling, we assume the following form
\cite{SPIN_ORBIT1,SPIN_ORBIT2,SPIN_ORBIT3}:
\begin{equation}
H_{\rm so} = i \sum_{j} \sum_{\alpha=x,y} c_j^{\dagger} 
{\mbox{\boldmath ${\bf \lambda}$}}^{(\alpha)}
\cdot {\mbox{\boldmath ${\bf \sigma}$}} ~c_{j+ a {\hat e}_{\alpha}} +
H.c.,
\label{eq_so}
\end{equation}
where $\mbox{\boldmath ${\bf \lambda}$}^{(\alpha)}=
\left( \lambda^{(\alpha)}_x, \lambda^{(\alpha)}_y \right)$,
${\hat e}_{\alpha}$ is a unit vector along
the $\alpha$-axis, and $a$ is the lattice constant.
Spin-orbit coupling (\ref{eq_so}) is produced by a buckling of 
the CuO$_2$ plane. An example is shown in Fig. \ref{fig_buckling}.
In the presence of spin-orbit coupling (\ref{eq_so}),
there is generally a Dzyaloshinskii-Moriya-type interaction
for the spin system. However, we ignore this term because it does
not play an important role in terms of the mechanism of interest.
In order to describe the spin system we introduce the Schwinger
bosons\cite{AROVAS}:
${\bf S}_j = \frac12 z^{\dagger}_j {\mbox{\boldmath ${\bf \sigma}$}}
z_j$ with $z_j^{\dagger}=\left( \begin{array}{cc} z_{j
\uparrow}^{\dagger} & z_{j\downarrow}^{\dagger} \end{array} \right)$
and $z_j = \left ( \begin{array}{cc} z_{j\uparrow} & z_{j\downarrow}
\end{array} \right)^T$. 
Here $z_{j\sigma}$ are boson fields and obey the constraint
$z_{j\uparrow}^{\dagger}z_{j\uparrow} + z_{j\downarrow}^{\dagger}
z_{j\downarrow} = 1$, since $S=1/2$.

We use the path-integral formulation to describe the system.
The action is given by
\begin{equation}
S=\int dt
\left\{ 	
       \sum_j \left[ \overline{c}_j (t)
       i \partial_t c_j(t) + 
       \overline{z}_j(t) i \partial_t z_j(t) \right]
       - H 
\right\}.
\end{equation}
The effect of the spin fluctuation on holes can be introduced by
performing a series of SU(2) transformations.
Let us focus on an $i-j$ bond along the $\alpha$-axis where $i$ and
$j$ are nearest-neighbor sites. We assume that the j-site belongs to
the A-sublattice and the i-site belongs to the B-sublattice.
The hopping term is given by
\begin{equation}
h_{ij}(\mbox{\boldmath ${\bf \lambda}$}^{(\alpha)}) = 
\sqrt{t_0^2+\lambda_{\alpha}^2} ~\overline{c}_i
\chi (\mbox{\boldmath ${\bf \lambda}$}^{(\alpha)}) c_j,
\end{equation}
where $\chi ( \mbox{\boldmath ${\bf \lambda}$}^{(\alpha)} ) =
\frac{1}{\sqrt{t_0^2+\lambda_{\alpha}^2}} \left( -t_0
\sigma_0 + i \mbox{\boldmath ${\bf \lambda}$}^{(\alpha)} \cdot
\mbox{\boldmath ${\bf \sigma}$} \right)$, with $\sigma_0$ the unit
matrix in spin space and $\lambda_{\alpha} =
|\mbox{\boldmath ${\bf \lambda}$}^{(\alpha)}|$.
Since $\lambda_{\alpha} \neq 0$, there is a twist in spin space.
Then, first we perform a transformation to diagonalize the hopping
term:
$c_j \rightarrow 
\chi^{\dagger} ( \mbox{\boldmath ${\bf \lambda}$}^{(\alpha)}) c_j$,
and $\overline{c}_i \rightarrow \overline{c}_i$.
By this transformation, we obtain 
$h_{ij} (\mbox{\boldmath ${\bf \lambda}$}^{(\alpha)}) 
= - \sqrt{t_0^2+\lambda_{\alpha}^2}
\overline{c}_i c_j$ up to $O((\lambda_{\alpha}/t_0)^2)$ in the phase
factor.
Second, we perform a transformation which diagonalizes the Kondo
coupling term: $c_j \rightarrow U_j c_j$, $\overline{c}_i \rightarrow
\overline{c}_i \left( i \sigma_y \overline{U}_i \right)$, 
where $U_l = \left( \begin{array}{cc}
z_{l \uparrow} & - \overline{z}_{l\downarrow} \\ z_{l\downarrow} &
\overline{z}_{l \uparrow} \end{array} \right)$ and $\overline{U}_l =
\left( \begin{array}{cc} \overline{z}_{l\uparrow} &
\overline{z}_{l\downarrow} \\ - z_{l \downarrow} & z_{l\uparrow} 
\end{array} \right)$.
By this transformation, SU(2) gauge fields which describe
the fluctuation of the spins are introduced\cite{DFM,MUDRY_FRADKIN}:
$
\overline{U}_i U_j 
\equiv
|\overline{U}_i U_j | \exp \left( -i a {\cal A}_{ji} \right).
$
The use of this transformation is based on the assumption that $J_K$
has a higher energy scale than any other parameter in the
Hamiltonian and the antiferromagnetic correlation between spins is
retained and its correlation length is much larger than the lattice
constant $a$.
In this transformation, the sign of Kondo coupling and the time
component of the gauge field: 
${\cal A}_t \equiv -i \overline{U} \partial_t U$ , for holes at
the B-sublattice is changed since there is an extra factor, 
$i \sigma_y$ for the B-sublattice.
For the amplitude fluctuation of the spins; $|\overline{U}_i U_j|$,
we take its mean value $\eta = \langle |\overline{U}_i U_j | \rangle$,
because it is a relatively higher energy mode than the phase
fluctuation of the spins.
Next, we perform the inverse transformation of the first transformation 
and we obtain 
\begin{eqnarray}
h_{ij} (\mbox{\boldmath ${\bf \lambda}$}^{(\alpha)}) 
&=&
- \sqrt{t_0^2+\lambda_{\alpha}^2} ~\eta ~\overline{c}_i (i \sigma_y)
\nonumber \\
& & \times \exp \left( - \frac{i}{t_0} \mbox{\boldmath ${\bf
\lambda}$}^{(\alpha)} \cdot
\mbox{\boldmath ${\bf \sigma}$} - i a {\cal A}_{ji} \right) c_j,
\label{eq_hopping}
\end{eqnarray}
up to $O((\lambda_{\alpha}/t_0)^2)$ in the exponent.

Compared with the model considered in ref.~\citen{MORINARI}, 
the derivation of the Chern-Simons term is complicated because we need 
to distinguish between the A- and B-sublattices.
However, there is a transformation which connects the model to that in 
ref.~\citen{MORINARI}, that is,
\begin{equation}
c_j \rightarrow c_j,
\hspace{2em} \overline{c}_i \rightarrow \overline{c}_i (- i\sigma_y).
\label{AF_F}
\end{equation}
After this transformation is performed, an additional sign change
occurs for both the Kondo coupling term and the time component
of the gauge field.
However, as we will discuss later, transformation (\ref{AF_F})
changes the symmetry of the pairing matrix. Therefore, we must
perform the inverse transformation to obtain the physical state of the 
original system.
In the following, we call the system obtained after transformation
(\ref{AF_F}) the ``F-system'' and the original system the ``AF-system.''
In the F-system, the action for the holes is given by
\begin{equation}
S_h= \int dt \sum_j \overline{c}_j(t) 
G^{-1} (\{ {\hat k}_{\mu} - {\cal A}_{\mu} \})
c_j (t),
\end{equation}
where ${\hat k}_{\mu}$ is defined by ${\rm e}^{ia{\hat k}_{\alpha}}c_j =
c_{j+a{\hat e}_{\alpha}}$, for $\mu=x,y$, 
and ${\hat k}_t = -i \partial_t$.
The inverse of the Green's function is given by
\begin{equation}
G^{-1} (\left\{ k_{\mu} \right\}) = (k_0 + 2 t_0 \eta
\sum_{\alpha =x,y} 
\cos k_{\alpha} )\sigma_0
- {\bf g} ({\bf k}) \cdot \mbox{\boldmath ${\bf \sigma}$},
\label{eq_Ginv_L}
\end{equation}
where
\begin{equation}
{\bf g} ({\bf k}) = \left(
2 \eta \sum_{\alpha=x,y} \lambda^{(\alpha)}_x \sin k_{\alpha},
2 \eta \sum_{\alpha=x,y} \lambda^{(\alpha)}_y \sin k_{\alpha},
-\frac{J_K}{4} \right).
\label{eq_g}
\end{equation}
Note that we cannot take the limit $J_K \rightarrow \infty$ in the
presence of spin-orbit coupling (\ref{eq_so}).
If we take the limit $J_K \rightarrow \infty$, then the spin of holes
is projected in the direction antiparallel to the spin at Cu sites. 
However, the hopping process always involves the opposite spin of
holes as long as $\lambda_{\alpha} \neq 0$. 

In order to calculate the Chern-Simons term, we take a continuum
limit.
(The condition of taking this limit will be discussed later.)
The induced Chern-Simons term is given by \cite{VOLOVIK,HST,READ_GREEN}
\begin{equation}
S_{\rm CS} = - \frac{\theta}{2\pi}
\int dt \int d^2 {\bf r}
{\cal A}_t^x \left( \partial_x {\cal A}^x_y - \partial_y {\cal A}^x_x
\right)
\label{eq_CS}
\end{equation}
where
\begin{equation}
\theta 
= \frac{1}{4\pi} \int d^2 {\bf k} 
\frac{{\bf g}({\bf k}) \cdot \left( \partial_{k_x} {\bf g}({\bf k}) 
\times \partial_{k_y} {\bf g}({\bf k}) \right)}{|{\bf g}({\bf
k})|^3}.
\label{eq_CSth}
\end{equation}
Note that only the Abelian Chern-Simons term appears in
eq. (\ref{eq_CS}) because the SU(2) gauge field ${\cal A}_{\mu}$ is
reduced to the Abelian Chern-Simons term upon using the curl-free
condition\cite{HST}.
We retain ${\cal A}_{\mu}^x$ because it describes the staggered
spin fluctuation.
From eq. (\ref{eq_CSth}), we obtain
\begin{equation}
\theta = - \frac12 \times {\rm sgn} \left( J_K \Lambda \right),
\label{eq_th}
\end{equation}
where $\Lambda \equiv \lambda_x^{(x)} \lambda_y^{(y)}-\lambda_y^{(x)}
\lambda_x^{(y)}$, and we have used the continuum form of (\ref{eq_g}): 
${\bf g}({\bf k}) = (2\eta
\sum_{\alpha}\lambda_x^{(\alpha)}k_{\alpha}, 2\eta
\sum_{\alpha}\lambda_y^{(\alpha)}k_{\alpha},$
$ -J_K/4)$.
In contrast to the anyon system\cite{ANYON1,ANYON2}, the value of
$\theta = \pm 1/2$ does not alter the statistics of
particles\cite{KIVELSON_ROKHSAR}.
We can extend the above calculation to a finite
temperature\cite{BDP}. However, if we concentrate on the region $k_B T
\ll J_K$, we can neglect finite temperature corrections.
Since spin-orbit coupling term (\ref{eq_so}) involves 
a process of hopping between different d-orbitals at the same site,
the external electromagnetic gauge field $A_{\mu}^{\rm ext}$ does not
couple to it.
Therefore, there is no Chern-Simons term for $A_{\mu}^{\rm ext}$.
In deriving eq. (\ref{eq_CS}), we have taken the continuum limit for
the gauge field ${\cal A}_{\mu}^x$. 
Since the length scale of the gauge field is given by 
$v/\Delta_{\rm sw}$,
where $\Delta_{\rm sw}$ and $v$ denote the gap and the velocity 
of the spin wave mode respectively,
the condition of taking the continuum limit is
$\Delta_{\rm sw}/(v/a) < \lambda_{\alpha}/t_0$, 
which can be seen from eq. (\ref{eq_hopping}).
For $\lambda_{\alpha}$, a rough estimation gives 
$\lambda_{\alpha} \sim 2~$meV\cite{SPIN_ORBIT1}.
We assume that this condition is satisfied in the underdoped region
because there $\Delta_{\rm sw}$ may be very small and $v/a$ is close
to the value of the undoped case $\sim 200~$meV\cite{SPIN_WAVE}.
Although the presence of spin-orbit couping is essential for the
derivation of the Chern-Simons term, it has no importance
for other physical processes. 
Therefore, we can neglect it in the following discussion.

Now we discuss the effect of the Chern-Simons term.
We take 
\begin{equation}
S=S_h+S_{\rm CS}+S_{\rm spin}, 
\end{equation}
for the effective action. The last term is the action for
the spin system and is given by the $CP_1$ model\cite{READ_SACHDEV}:
$S_{\rm spin} = (2/g)
\int d^3 x \sum_{\sigma}
\left[ 
\left| \left( \partial_{\mu} - i {\cal A}_{\mu}^x \right) 
z_{\sigma} \right|^2 
+ \left( \Delta_{\rm sw}^2/ v^2 \right) |z_{\sigma}|^2
\right]$.
By integrating out the ${\cal A}_t^x$, we obtain the relationship
between the spin density of the hole and the ``magnetic'' field
${\cal B} ({\bf r},t) = \partial_x {\cal A}_y^x ({\bf r},t) -
\partial_y {\cal A}_x^x ({\bf r},t)$.
If we take the x-axis as the quantization axis for the spin, 
we obtain
$\sum_{\sigma} s_{\sigma} \rho_{\sigma} ({\bf r},t) =
\frac{\theta}{2\pi} {\cal B}({\bf r},t)$,
where $s_{\sigma}=1$ for $\sigma=\uparrow$ and 
$s_{\sigma}=-1$ for $\sigma=\downarrow$. 
In the AF-system, this relation involves the isospin index, that is, 
\begin{equation}
\sum_{\sigma} s_{\sigma} s_{\tau} \rho_{\sigma} ({\bf r},t)
= \frac{\theta}{2\pi} {\cal B}({\bf r},t).
\label{eq_flux_AF}
\end{equation}
Here $s_{\tau}=1(-1)$ for ${\bf r}$ belongs to the A(B)-sublattice.
Therefore, 
$\uparrow$($\downarrow$)-spin at the A-sublattice induces a
(anti-)skyrmion excitation in the localized spin system and
$\downarrow$($\uparrow$)-spin at the B-sublattice induces a
(anti-)skyrmion excitation in the localized spin system.
Since the skyrmion and anti-skyrmion excitations introduce disorder
into the localized spin system\cite{BELAVIN_POLYAKOV}
and the number of them is the same as that of holes,
disorder in the spin system increases upon doping.
If the hole density is sufficiently small that the magnetic long-range
order is preserved, then the Meissner effect occurs for the gauge field
${\cal A}^x_{\mu}$ and 
holes are pinned because skyrmions break the translational invariance,
that is, the system is an insulator.

After the magnetic long-range order is destroyed by the skyrmion
excitations\cite{SKYRMION5,MARINO_NETO}, 
the Chern-Simons term becomes dominant in
the long wavelength and low-energy physics.
For the holes, it leads to a pairing state.
Coupling between the hole current and the gauge
field ${\cal A}^x_{\mu}$, in the F-system is given by
\begin{equation}
S_{j-{\cal A}} = \sum_{\sigma=\uparrow,\downarrow} 
\int dt \int d^2 {\bf r}~s_{\sigma}
{\bf j}_{\sigma} ({\bf r},t) \cdot 
\mbox{\boldmath ${\bf {\cal A}}$}^x ({\bf r},t),
\label{eq_j_A}
\end{equation}
where $j_{\sigma} ({\bf r},t)$ is the hole current for
$\sigma$-spin. 
Since eq. (\ref{eq_j_A}) describes minimal coupling between the
hole current and the gauge field $\mbox{\boldmath ${\bf {\cal
A}}$}^x$, it gives rise to a Lorentz force.
Such a Lorentz force is induced between holes
passing each other.
Therefore, it leads to a chiral pairing state. The chirality is
determined by the sign of $\theta$.
From its pairing mechanism, the possibility of the s-wave pairing state
is excluded.

Now we investigate the pairing state of the AF-system through the
F-system. Before doing that, we must know the relationship of the
pairing matrix between them.
We assume that $i$ and $j$ are nearest neighbor sites.
If we take $\Delta_{ij}^s 
= \langle c_{i\uparrow} c_{j \downarrow}
- c_{i\downarrow} c_{j \uparrow} \rangle$ for the spin-singlet pairing
order parameter for the AF-system, then after performing
transformation (\ref{AF_F}) we obtain 
$\Delta_{ij}^s 
\rightarrow \langle c_{i\uparrow} c_{j \uparrow}
+ c_{i\downarrow} c_{j \downarrow} \rangle$.
Therefore, the spin-singlet pairing state is transformed
into the spin-triplet pairing state and vice versa.
In ${\bf k}$-space, holes at the A-sublattice are described by the
fields $\alpha_{{\bf k}\sigma}=\frac{1}{\sqrt{2}} 
\left( c_{{\bf k}\sigma} + c_{{\bf k}+{\bf Q} \sigma} \right)$
and holes at the B-sublattice are described by
$\beta_{{\bf k}\sigma}=\frac{1}{\sqrt{2}} 
\left( c_{{\bf k}\sigma} - c_{{\bf k}+{\bf Q} \sigma} \right)$,
where ${\bf Q}=(\pi/a,\pi/a)$.
Here, we assume that the A-sublattice is the set of 
$({\rm even},{\rm even})$ and $({\rm odd},{\rm odd})$ and the
B-sublattice is the set of 
$({\rm even},{\rm odd})$ and $({\rm odd},{\rm even})$.
The pairing matrix may be given by
\begin{equation}
\Delta^{\bf k}_{\sigma_1 \sigma_2} 
= 
{\sum_{{\bf k}^{\prime}}}^{\prime} 
V_{{\bf k} {\bf k}^{\prime}}^{\rm AF} 
\langle \beta_{-{\bf k}^{\prime} \sigma_2} \alpha_{{\bf k}^{\prime}
\sigma_1} \rangle_{\rm AF},
\label{eq_gapAF}
\end{equation}
where $\sum_{\bf k}^{\prime} (f_{\bf k}+f_{{\bf k}+{\bf Q}}) =
\sum_{\bf k} f_{\bf k}$.
Here, we do not need the explicit form of 
$V_{{\bf k} {\bf k}^{\prime}}^{\rm AF}$, 
because we solve the gap equation through that of the F-system.
Transformation (\ref{AF_F}) in ${\bf k}$-space is given by
$\beta_{{\bf k}\sigma} \rightarrow i \sigma_y \beta_{{\bf k}\sigma}$.
By performing this transformation, the gap equation (\ref{eq_gapAF})
is transformed into
\begin{equation}
\Delta^{\bf k}_{\sigma_1 \sigma_2} 
= 
{\sum_{{\bf k}^{\prime}}}^{\prime} 
V_{{\bf k} {\bf k}^{\prime}}^{\rm F} 
\langle \left( i\sigma_y \beta_{-{\bf k}^{\prime}} \right)_{\sigma_2}
\alpha_{{\bf k}^{\prime} \sigma_1} \rangle_{\rm F}.
\label{eq_gap_F}
\end{equation}
For the singlet pairing case 
$\Delta_{\uparrow \downarrow}^{\bf k}= -\Delta_{\downarrow
\uparrow}^{\bf k}$, eq. (\ref{eq_gap_F}) is reduced to
\begin{equation}
\Delta_{\uparrow \downarrow}^{\bf k} = 
\left( \Delta_{\uparrow \uparrow}^{(1){\bf k}}
+ \Delta_{\uparrow \uparrow}^{(2){\bf k}} \right)/2, 
\label{eq_Delta}
\end{equation}
where
\begin{eqnarray}
\Delta_{\uparrow \uparrow}^{(1){\bf k}}
&=&
{\sum_{{\bf k}^{\prime}}}^{\prime} 
V_{{\bf k}{\bf k}^{\prime}}^{\rm F} 
\langle c_{-{\bf k}^{\prime} \uparrow} c_{{\bf k}^{\prime} \uparrow}
\rangle_{\rm F},
\label{eq_Delta1} \\
\Delta_{\uparrow \uparrow}^{(2){\bf k}}
&=& -
{\sum_{{\bf k}^{\prime}}}^{\prime} 
V_{{\bf k}{\bf k}^{\prime}}^{\rm F} 
\langle c_{-{\bf k}^{\prime}+{\bf Q} \uparrow} c_{{\bf k}^{\prime}
+ {\bf Q} \uparrow} \rangle_{\rm F}.
\label{eq_Delta2}
\end{eqnarray}
The minus sign in eq. (\ref{eq_Delta2}) originates from the sign
change in the kinetic term in $G^{-1} \left( k_0, {\bf k}+{\bf Q}
\right)$.
Although the vector ${\bf g}({\bf k})$ changes as
${\bf g} \left( {\bf k}+{\bf Q} \right) = {\rm diag}(-1,-1,1) 
{\bf g} \left( {\bf k} \right)$, the sign of $\theta$ does not change.
In the continuum approximation, eqs. (\ref{eq_Delta1}) and
(\ref{eq_Delta2}) are reduced to
\begin{equation}
\Delta_{\bf k} = \pm \sum_{{\bf k}^{\prime}} 
\frac{4\pi i}{\theta} 
\frac{{\bf k}\times {\bf k}^{\prime}}{|{\bf k}-{\bf k}^{\prime}|^2}
\frac{\Delta_{\bf k}}{2E_{\bf k}},
\label{eq_gap}
\end{equation}
where $E_{\bf k} = \sqrt{\xi_{\bf k}^2 + |\Delta_{\bf k}|^2}$, 
at $T=0$.
Following the analysis of ref.~\citen{GWW}, we can solve
eq. (\ref{eq_gap}). The solution is given by
$
\Delta_{\uparrow \uparrow}^{(1){\bf k}} = \Delta_k \exp \left( \pm
2i\ell \theta_{\bf k} \right)$,
and $\Delta_{\uparrow \uparrow}^{(2){\bf k}} = \Delta_k \exp
\left( \mp 2i\ell \theta_{\bf k} \right)$,
where $\Delta_k$ is a function of $k=|{\bf k}|$ 
and $\theta_{\bf k}=\arctan k_y/k_x$.
Here, $2\ell$ is the relative angular momentum of the Cooper pair and
is not equal to zero.
The gap of superconductivity $\Delta_{k_F}$ is of the order of
Fermi energy $\epsilon_F$ and
the cohrence length of superconductivity $\xi_{\rm SC}$ 
is given by
$\xi_{\rm SC}/a \sim \frac{\epsilon_F}{4\sqrt{x}\Delta_{k_F}}$.
Here, $x$ is the hole concentration.
Since $\Delta_{k_F}$ is of the order of $\epsilon_F$,
this value may be smaller than 
the average distance between holes: $\sim a/\sqrt{x}$.
From eq. (\ref{eq_Delta}), we obtain
\begin{equation}
\Delta_{\uparrow \downarrow}^{\bf k} = - \Delta_{\downarrow
\uparrow}^{\bf k} = \Delta_k \cos \left( 2\ell \theta_{\bf k} \right).
\label{eq_singlet}
\end{equation}
Since the smallest $\ell$ is realized in the ground state,
we set $\ell=1$.
In this case, eq. (\ref{eq_singlet}) describes the $d_{x^2-y^2}$
pairing state because $\cos \left( 2\theta_{\bf k} \right) = \left
( k_x^2 - k_y^2 \right)/k^2$.
For the triplet pairing case, we find that
$\Delta_{\uparrow \uparrow}^{\bf k} =
\Delta_{\downarrow \downarrow}^{\bf k} = 
\left( 
\Delta_{\uparrow \downarrow}^{(1){\bf k}} + 
\Delta_{\uparrow \downarrow}^{(2){\bf k}}
\right)/2$, and
$\Delta_{\uparrow \downarrow}^{\bf k}
=\Delta_{\downarrow \uparrow}^{\bf k}=0$.
However, such a pairing state is not stable in the bulk of the system
because the d-vector \cite{LEGGETT} satisfies ${\bf d}_{\bf k}
\parallel {\hat e}_y$, that is, the spins of
Cooper pairs lie in the plane perpendicular to the y-axis.
As a result, the pairing state has spin-singlet and $d_{x^2-y^2}$
symmetry.

There is also some contribution to the pairing mechanism from other
spin fluctuations, which may be characterized by the Maxwell
term: 
$\sim - \frac14 \left ( \partial_{\mu} {\cal A}^x_{\nu} - \partial_{\nu}
{\cal A}^x_{\mu} \right)^2$
in the gauge field description because coupling to the spin system 
is mediated by the gauge field ${\cal A}_{\mu}^x$.
Meanwhile, our spin fluctuation is characterized by the Chern-Simons
term. Since there is an extra derivative for the former compared with
the latter, our mechanism may be more dominant in the long wavelength
and the low-energy limit than other spin fluctuation mechanisms.
Moreover, the Chern-Simons term only exists in the $2+1$ dimension.
Therefore, our spin fluctuation is unique to the $2+1$ dimension. 
In contrast, the Maxwell term exists in any dimension.

For the application to the orthorhombic phase of
La$_{2-x}$Sr$_x$CuO$_4$\cite{SPIN_ORBIT1,SPIN_ORBIT2,SPIN_ORBIT3}, 
we require one more transformation after
transformation (\ref{AF_F}), that is,
$c_{({\rm odd},{\rm even})} \rightarrow 
i\sigma_z c_{({\rm odd},{\rm even})}$, and
$c_{({\rm even},{\rm odd})} \rightarrow 
-i\sigma_z c_{({\rm even},{\rm odd})}$.
Also in this case, the value of $\theta$ is given by eq. (\ref{eq_th})
and the pairing state has spin-singlet and $d_{x^2-y^2}$ symmetry.

In summary, we have studied a model of the CuO$_2$ plane with buckling
and have shown that the Chern-Simons term for the gauge field, which
describes the fluctuation of the spin system, is induced.
Through this Chern-Simons term, the doped hole behaves like a skyrmion
or anti-skyrmion excitation depending on its spin or isospin, that
is, whether it resides on the A-sublattice or B-sublattice.
After the antiferromagnetic long-range order is destroyed by the
skyrmion excitations, the Chern-Simons term becomes dominant for
long wavelength and low-energy physics and leads to the spin-singlet
$d_{x^2-y^2}$ superconducting state.

The author would like to thank M. Sigrist, K. Ohgushi, J. Goryo,
A. Furusaki, and K. K. Ng for helpful discussions. 
This work was supported by a Grant-in-Aid from the Ministry
of Education, Culture, Sports, Science and Technology.

\end{document}